\documentclass[10pt,conference]{IEEEtran}
\IEEEoverridecommandlockouts
\usepackage{cite}
\usepackage{amsmath,amssymb,amsfonts}
\usepackage{algorithmic}
\usepackage{graphicx}
\usepackage{textcomp}
\usepackage{listings}
\usepackage{xcolor}
\usepackage{url}
\usepackage{hyperref}
\usepackage{booktabs}
\usepackage{colortbl}
\usepackage{threeparttable}
\usepackage{multirow} 
\hypersetup{
colorlinks=true,
linkcolor=black,
citecolor=black,
urlcolor=blue
}

\def\BibTeX{{\rm B\kern-.05em{\sc i\kern-.025em b}\kern-.08em
    T\kern-.1667em\lower.7ex\hbox{E}\kern-.125emX}}
\begin{document}

\title{ScalerEval: Automated and Consistent Evaluation Testbed for Auto-scalers in Microservices}

\author{
	\IEEEauthorblockN{
		Shuaiyu Xie\IEEEauthorrefmark{1},
        Jian Wang\thanks{† Jian Wang and Bing Li are the corresponding authors.}\IEEEauthorrefmark{1}\IEEEauthorrefmark{2}, 
		Yang Luo\IEEEauthorrefmark{1}, 
        Yunqing Yong\IEEEauthorrefmark{1},
        Yuzhen Tan\IEEEauthorrefmark{1},
		and Bing Li\IEEEauthorrefmark{1}\IEEEauthorrefmark{2}} 
	\IEEEauthorblockA{\IEEEauthorrefmark{1}School of Computer Science, Wuhan University, China}
    \IEEEauthorblockA{\IEEEauthorrefmark{2}Zhongguancun Laboratory, China}
    \IEEEauthorblockA{Email: \{theory, jianwang, sanqine, yongyunqing, tanyuzhen, bingli\}@whu.edu.cn}
}


\maketitle

\begin{abstract}
Auto-scaling is an automated approach that dynamically provisions resources for microservices to accommodate fluctuating workloads. Despite the introduction of many sophisticated auto-scaling algorithms, evaluating auto-scalers remains time-consuming and labor-intensive, as it requires the implementation of numerous fundamental interfaces, complex manual operations, and in-depth domain knowledge. Besides, frequent human intervention can inevitably introduce operational errors, leading to inconsistencies in the evaluation of different auto-scalers. To address these issues, we present ScalerEval, an end-to-end automated and consistent testbed for auto-scalers in microservices. ScalerEval integrates essential fundamental interfaces for implementation of auto-scalers and further orchestrates a one-click evaluation workflow for researchers. The source code is publicly available at \href{https://github.com/WHU-AISE/ScalerEval}{https://github.com/WHU-AISE/ScalerEval}.
\end{abstract}


\section{Introduction}
The widespread adoption of microservice architecture has sparked a paradigm shift in cloud-native applications. This prominent architecture decomposes monolithic software into fine-grained and loosely coupled service units, aka microservices, enabling rapid iteration and flexible scalability for system development. Typically, a microservice instantiates several identical replicas, which run in process-isolated containers and share the external workloads. Despite many benefits of microservice architecture, microservices often face performance challenges arising from unpredicable workloads and insufficient replicas. Although cloud providers can provision excessive replicas to mitigate performance degradation, this redundancy mechanism leads to significant resource waste and financial loss.

Auto-scaling is an automated technique designed to elastically adjust the number of replicas for each microservice according to workload variations, ensuring the service level agreement (SLA) and reducing redundant resources. The evaluation process of an auto-scaling algorithm typically consists of five steps: benchmark initialization, scaler registration, workload injection, metric collection, and performance assessment \cite{xie2024pbscaler, luo2022erms}. Unfortunately, existing studies primarily focus on refining auto-scaling algorithms while neglecting the standardization and automation of the evaluation process, resulting in the following challenges:

\begin{enumerate}
    \item \textbf{Tedious efforts in building experimental infrastructure}. The implementation of auto-scaling requires developing various fundamental functions, such as observability, scaling operations, and workload injection, which serve as prerequisites for auto-scalers. However, designing these interfaces from scratch is a complex and time-consuming task, even for experienced engineers. Consequently, it is imperative to devise a valid and comprehensive infrastructure to facilitate the implementation of auto-scalers.
    
    \item \textbf{Complexity and error-proneness of manual operation}. Auto-scaling evaluation often involves numerous manual operations across different machines, such as launching benchmarks, registering scalers, and injecting workloads. However, introducing extensive human intervention is both laborious and error-prone. To make matters worse, inadvertent discrepancies caused by manual operations can lead to inconsistencies in the evaluation process, thereby undermining the validity and reproducibility of experimental results.
    
    \item \textbf{Inconsistent benchmark across evaluations}. The runtime state of microservices is in constant flux because of some changes occurring in environment, cache, or persistent storage. This unstable run-time behavior furnishes inconsistent benchmarks for each evaluation process, resulting in unfairness across different auto-scalers. To mitigate the inconsistency problem, the selected benchmark and the corresponding infrastructure should be reset before auto-scaling experiments.

\end{enumerate}

To address these challenges, we introduce ScalerEval, an end-to-end cohesive testbed that automates the key stages of auto-scaling evaluation, including benchmark initialization, scaler registration, workload injection, metric collection, and performance assessment. Considering the inherent complexity of manual operations (e.g., workload injection and scaler registration), ScalerEval weaves and automates these operations into an execution workflow, eliminating biases caused by human errors. Regarding the integration of various custom auto-scalers, we define a scaler template and further provide sufficient fundamental interfaces, including observability and operability for target systems. To ensure consistent benchmarks across evaluations, ScalerEval proactively resets microservice-based systems and monitoring infrastructure before each test. The major contributions of ScalerEval are as follows:

\begin{itemize}
    \item We present an automated and consistent testbed for auto-scalers in microservices, namely ScalerEval, which covers integral components in auto-scaling, including benchmark initialization, scaler registration, workload injection, metric collection, and performance assessment. 
    
    
    \item We provide comprehensive interfaces to facilitate the implementation of auto-scalers. We also implemented and evaluated several state-of-the-art auto-scalers on two widely-used microservice-based systems. 

    \item We open source ScalerEval \cite{ScalerEval} to support researchers in efficiently advancing their exploration of auto-scaling.
\end{itemize}

\section{Background \& Related Work}
\subsection{Auto-scaling for Microservices}
Microservice-based systems often suffer from performance issues due to volatile workloads and inadequate resources. To meet the service level agreement (SLA), cloud providers tend to over-provision replicas for microservices, resulting in excessive resource consumption \cite{qiu2020firm}. As such, significant efforts have been made to investigate the autoscaling mechanism, thereby allocating replicas to align with demands. Current auto-scalers can be categorized into three types according to the scaling policy. (1) \textbf{Threshold-based approach}. Most cloud providers, such as AWS \cite{AWS} and Alibaba Cloud \cite{aliyun}, offer an efficient implementation of auto-scalers based on static resource thresholds. However, a uniform, coarse-grained threshold can hardly serve as a silver bullet for all microservices. (2) \textbf{Control theory-based approach}. By setting a target value (e.g., latency), some methods \cite{baarzi2021showar} continuously adjust decisions based on system feedback, driving the system state to approach the desired state. (3) \textbf{Model-based approach}. The premise of these auto-scalers is to build a fitting model to profile some system behaviors, including workload prediction and latency estimation. The underlying rationales of this model can be based on queuing theory \cite{chen2024derm}, machine learning \cite{qiu2020firm, xie2024pbscaler}, or deep learning \cite{bai2024DRPC, park2021graf}. Under the guidance of trained models, auto-scalers can preemptively optimize the replicas provision on demand.

\subsection{Auto-scaling Testbed}
Because of the black-box nature of the relationship between microservice performance (e.g., latency), workload, and allocated resources, evaluating auto-scalers solely based on offline static datasets is hardly convincing. Hence, most studies assess the performance of auto-scalers on several online microservice benchmarks. A comprehensive evaluation testbed for auto-scalers involves five steps: (1) \textbf{Benchmark initialization}. This step aims to start real microservice-based systems with the corresponding infrastructure, including monitoring systems and ingress gateways. (2) \textbf{Scaler registration}. Typically, auto-scalers should be registered as independent processes, interacting with monitoring systems and executing scaling actions according to workloads. (3) \textbf{Workload injection}. To better align with production scenarios, this step generates user threads based on real-world patterns, thereby sending test traffic to microservice-based systems. Note that the load-generating process is typically deployed outside the cluster, avoiding resource contention with the microservice-based system. (4) \textbf{Metric collection}. After scaling experiments, we need to collect sufficient metrics during the test period, such as tail latency, allocated replicas, and resource consumption, ensuring comprehensive observability for the test system. (5) \textbf{Performance assessment}. Considering the trade-off between performance and costs, the assessment of auto-scalers highlights the SLA violation rates and resource consumption (i.e., CPU and memory). Although previous works have implemented some of these individual steps, they have yet to integrate them into an automated one-click workflow, thereby increasing experimental complexity and hindering progress in the field.

\begin{figure}
    \centering
    \includegraphics[width=\linewidth]{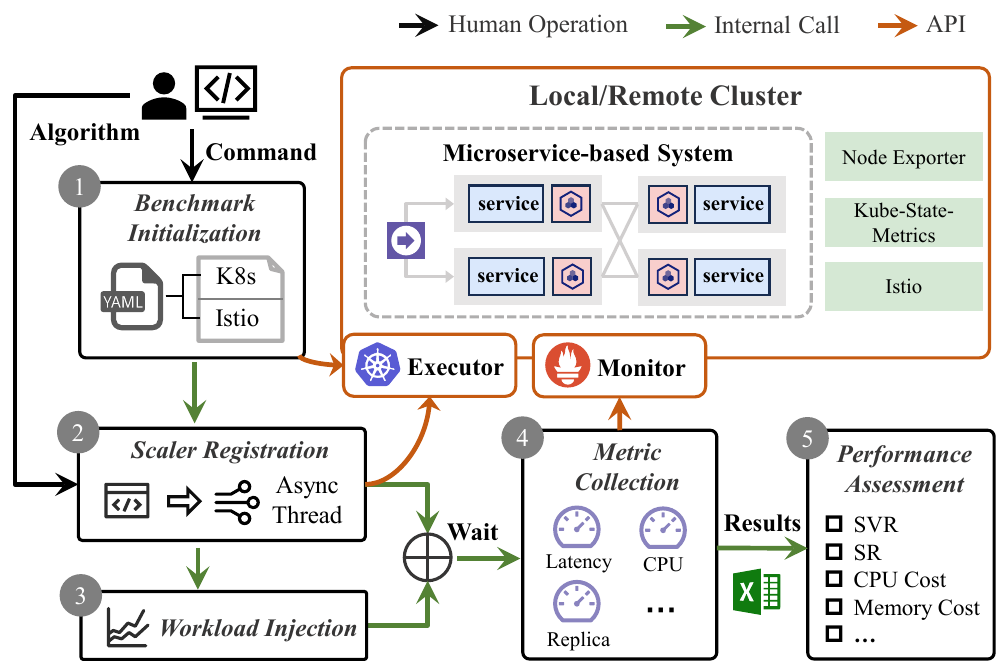}
    \caption{Overview of ScalerEval.}
    \vspace{-4mm}
    \label{fig:structure}
\end{figure}

\section{System Design}
ScalerEval is an automated and consistent testbed designed to provide an one-click evaluation workflow for auto-scalers. Fig. \ref{fig:structure} illustrates the whole ScalerEval process, encompassing five typical stages used in auto-scaling evaluation. ScalerEval automatically weaves these isolated stages into a complete evaluation lifecycle, which can be executed directly upon a single command.

\subsection{Benchmark Initialization}
In this stage, we start the microservice-based system under test and manage the traffic. Note that we clear any residual system state before this stage, eliminating the adverse effects caused by cache or persistent storage. This proactive reset operation ensures the consistency and independence of every auto-scaling evaluation.

\textbf{System startup}. We choose containerized microservice-based systems as benchmarks and use Kubernetes \cite{K8s} for container orchestration. For each benchmark, users need to prepare an object configuration file in YAML format, which records the Kubernetes resources required by the benchmark, including services and deployments. At the beginning of experiments, ScalerEval automatically scans the specified benchmark's object configuration file, thereby creating the namespace and corresponding Kubernetes resources.



\textbf{Traffic management}. To direct external traffic to microservice-based systems, ScalerEval provides a manifest file for traffic management based on Istio \cite{Istio}. This file configures an ingress gateway to filter traffic and defines the routing rules for the entry microservice. Users can easily customize the traffic rules in this file, which will be automatically loaded after the system starts.

\begin{figure}
    \centering
    \includegraphics[width=1\linewidth]{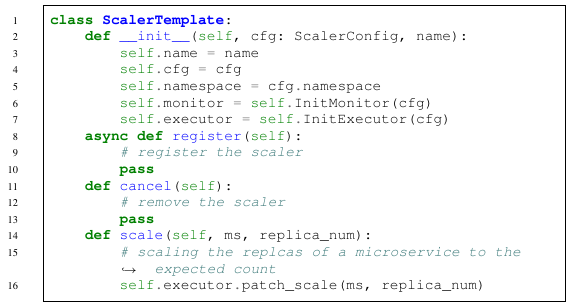}
    \caption{Template of auto-scalers.}
    \label{fig:template}
\end{figure}

\subsection{Scaler Registration}
Once the benchmark is initialized, ScalerEval subsequently registers an auto-scaler to perform scaling actions on demand. We also furnish a basic infrastructure to support the implementation of different auto-scalers.

\textbf{Fundamental interfaces}. We consider most auto-scalers to follow the MAPE-K architecture \cite{arcaini2015modeling, ahmad2024smart, bai2024DRPC}, which consists of Monitor, Analyzer, Planner, Executor, and a Knowledge Base. we provide fundamental implementations for the Monitor and Executor, as these two components are similar across different auto-scalers. Specifically, ScalerEval provides basic functions for monitoring system metrics based on Prometheus \cite{prometheus}, including latency, workloads, and resource usage. Regarding Executor, we implement common operations for Kubernetes resources based on the Kubernetes API. By integrating the Monitor and Executor, auto-scalers can readily perceive the system behavior and modify the Kubernetes resources, faciliting the development of personalized auto-scalers.

\textbf{Scaler template}. We define a universal template to enable seamless integration of custom auto-scalers into ScalerEval. Fig. \ref{fig:template} shows the auto-scaler template, which defines the lifecycle of auto-scalers through three functions: \textit{register}, \textit{cancel}, and \textit{scale}. ScalerEval invokes the \textit{register} function to start the auto-scaler after the benchmark is launched and call the \textit{cancel} function to remove it at the end of the experiment. ScalerEval allows users to design and extend auto-scalers with specific logic, as long as they adhere to the template specification.

\textbf{Asynchronous registration}. ScalerEval registers an asynchronous thread to execute the auto-scaler, preventing it from blocking the main process. Note that this thread runs outside the cluster and does not consume cluster resources. It interacts with the microservice-based system inside the cluster through the Monitor and Executor.

\subsection{Workload Injection}
Parallel to the scaler registration module, ScalerEval uses Locust \cite{Locust} to generate external traffic for load testing. In particular, ScalerEval allows users to customize the load shape based on a CSV file. For example, users can simulate load spikes by adjusting user counts for specific times in the CSV file. After the workload injection, ScalerEval calls the \textit{cancel} function to signal the auto-scaler, terminating the corresponding asynchronous thread and releasing resources.

\begin{table}[t]
  \centering
  \caption{Metrics collected from Prometheus.}
    \begin{tabular}{c}
    \toprule
    \rowcolor[rgb]{ .906,  .902,  .902} \textbf{Node-level Metric} \\
    \midrule
    node\_cpu\_seconds\_total, node\_memory\_MemFree\_bytes \\
    node\_memory\_MemTotal\_bytes, node\_memory\_Buffers\_bytes \\
    \midrule
    \rowcolor[rgb]{ .906,  .902,  .902} \textbf{Service-level Metric} \\
    \midrule
    kube\_pod\_info, istio\_requests\_total,  \\
    container\_cpu\_usage\_seconds\_total, container\_memory\_usage\_bytes \\
    istio\_request\_duration\_milliseconds\_bucket,  \\
    \bottomrule
    \end{tabular}%
  \label{tab:metrics}%
  \vspace{-4mm}
\end{table}%

\subsection{Metric Collection and Performance Assessment}
ScalerEval starts the metric collection process as soon as the workload injection is complete. Table. \ref{tab:metrics} depicts the metric types collected from Prometheus. Specifically, we export various metrics at the service and node levels during the workload injection period, including resource usage, request duration, and request counts. All metrics are recorded at fixed intervals and stored in a CSV file, reflecting the performance of a microservice-based system during this period.

To assess the performance and cost of auto-scalers, ScalerEval introduces several typical indicators: SLA violation rate (SVR), success rate (SR), and total CPU and memory usage. In addition, we provide basic visualizations for the collected metrics, facilitating further analysis of the results.

\section{Evaluation}
In this section, we conducted auto-scaling experiments based on ScalerEval, demonstrating its actionability and scalability. We conducted auto-scaling experiments in a local cluster encompassing one master node and one worker node, with a total of 32 CPU cores and 96 GB of memory. Specifically, this cluster needs to be pre-equipped with Kubernetes \cite{K8s}, Istio \cite{Istio}, Node Exporter \cite{NodeExporter}, and Kube-State-Metrics \cite{kubeStateMetric}, enabling essential microservice-oriented management capabilities, including container orchestration, traffic routing, and observability.

\begin{table*}[t]
  \centering
  \caption{Performance comparison of auto-scalers.}
    \begin{threeparttable}
  \resizebox{\linewidth}{!}{
    \begin{tabular}{c|cccc|cccc}
    \toprule
    \multirow{2}[4]{*}{Approach} & \multicolumn{4}{c|}{Online Boutique} & \multicolumn{4}{c}{SockShop} \\
\cmidrule{2-9}          & SVR (\%) & SR(\%) & CPU ($\times 10^2$ Cores) \tnote{\dag} & Memory ($\times 10^3$ MB)\tnote{\dag}  & SVR (\%) & SR(\%) & CPU ($\times 10^2$ Cores) & Memory ($\times 10^3$ MB) \\
    \midrule
    NoneScaler & 0.96  & 0.96  & 6.23  & 10.39  & 0.99  & 0.88  & 9.20  & 47.54  \\
    KHPA-20 & 0.03  & 1.00  & 20.95  & 39.74  & 0.04  & 0.96  & 33.60  & 143.14  \\
    KHPA-50 & 0.03  & 1.00  & 19.14  & 22.77  & 0.05  & 0.96  & 38.54  & 154.54  \\
    KHPA-80 & 0.11  & 1.00  & 18.39  & 21.30  & 0.04  & 0.96  & 28.83  & 96.32  \\
    Showar & 0.02  & 1.00  & 20.96  & 42.46  & 0.07  & 0.94  & 46.51  & 235.96  \\
    PBScaler & 0.03  & 1.00  & 17.89  & 24.04  & 0.04  & 0.97  & 31.29  & 97.05  \\
    \bottomrule
    \end{tabular}}%
        \begin{tablenotes}
            \footnotesize
            \item[\dag] \textit{We summed the CPU and memory usage per second to obtain the results presented in the table.}
        \end{tablenotes}
    \end{threeparttable}
  \label{tab:performance}%
  \vspace{-4mm}
\end{table*}%

\subsection{Experimental Setup}
\textbf{Baseline Methods}. We reproduced and evaluated three state-of-the-art horizontal auto-scalers from academia and industry: PBScaler \cite{xie2024pbscaler}, Showar \cite{baarzi2021showar}, and Kubernetes horizontal pod auto-scaler (KHPA) \cite{KHPA}. For KHPA, we derived three variants that use CPU usage thresholds of 20\%, 50\%, and 80\%, respectively. To manifest the effectiveness of auto-scalers, we also introduced a variant called NoneScaler, which evaluates the system performance without applying any scaling operations.

\textbf{Benchmarks and Workloads}. We selected two popular microservice applications: Online Boutique \cite{hipster} and Sock Shop \cite{sockshop}, as evaluation benchmarks. To align with real-world scenarios, we injected 20 minutes of dynamic workloads derived from the Wiki-Pageviews dataset recorded on March 16, 2015. Considering the resource constraints of the cluster, we scaled the number of users, capping the maximum concurrent user counts at 740.


\subsection{Preliminary Experiments}
We evaluate selected auto-scalers on ScalerEval and report their indicators, including SLA violation rates (SVR), success rates (SR), and total resource usage. For the SVR, we configure the SLA as 500 ms. Table \ref{tab:performance} summarized a comparison of these auto-scalers across these indicators. Considering the system performance, all auto-scalers significantly decrease the SVR by at least 0.85 compared to NoneScaler which does not perform any scaling actions. This result demonstrates the effectiveness of auto-scalers implemented based on ScalerEval. Regarding the Online Boutique, Showar yields a slightly lower SVR than other baselines, owing to its agile control algorithm. However, its marginal advantage comes at the cost of high resource consumption. Moreover, the SVR and resource consumption of the KHPA algorithm show a nearly inverse correlation as the CPU usage threshold increases. This is because a higher CPU threshold signifies a more conservative scaling mechanism. The validation of these patterns, on the other hand, further proves the effectiveness of ScalerEval as an auto-scaling testbed.

\section{Conclusion}
In this paper, we present an end-to-end automated and consistent testbed, ScalerEval, which integrates a one-click evaluation workflow for auto-scaling experiments. To facilitate the integration of different auto-scalers, ScalerEval provides sufficient fundamental interfaces, ensuring observability and operability for microservice-based systems. To the best of our knowledge, this is the first automated testbed that covers the auto-scaling lifecycle. In the future, we plan to provide infrastructure support for vertical auto-scalers, enabling fine-grained resource allocation for containers. Furthermore, we will explore the integration of additional tasks, such as microservice deployment and traffic scheduling.

\bibliographystyle{IEEEtran}
\bibliography{scalerEval}
\end{document}